%% file: 0_main.tex
\documentclass{article}
\usepackage{array}
\usepackage{spconf,amsmath,graphicx}
\usepackage{amsfonts}
\usepackage{booktabs}
\usepackage{blindtext}
\usepackage{hyperref}
\usepackage{multirow}
\usepackage[dvipsnames]{xcolor}
\usepackage{tabularx}

\newcommand{\RN}[1]{%
  \textup{\uppercase\expandafter{\romannumeral#1}}%
}

\title{VarianceFlow: High-quality and Controllable Text-to-Speech \newline Using Variance Information via Normalizing Flow}
\name{Yoonhyung Lee\textsuperscript{1*}\thanks{*This work is done during an internship program at NCSOFT.} \qquad Jinhyeok Yang\textsuperscript{2} \qquad Kyomin Jung\textsuperscript{1}}
\address{
\textsuperscript{1}Seoul National University, Dept. of Electrical and Computer Engineering, Republic of Korea\\
\textsuperscript{2}Speech AI Lab, NCSOFT, Republic of Korea
}

\begin{document}
%
\maketitle
\begin{abstract}
\input{1_abstract.tex}
\end{abstract}
\vspace{-0.2cm}
\input{2_introduction}
\input{4_methodology}
\input{5_experiments}
\input{6_conclusion}
\input{7_acknowledge}

\bibliographystyle{IEEEbib}
\bibliography{references}

\end{document}

%% file: 1_abstract.tex
There are two types of methods for non-autoregressive text-to-speech models to learn the one-to-many relationship between text and speech effectively.
The first one is to use an advanced generative framework such as normalizing flow (NF).
The second one is to use variance information such as pitch or energy together when generating speech.
For the second type, it is also possible to control the variance factors by adjusting the variance values provided to a model.
In this paper, we propose a novel model called VarianceFlow combining the advantages of the two types.
By modeling the variance with NF, VarianceFlow predicts the variance information more precisely with improved speech quality.
Also, the objective function of NF makes the model use the variance information and the text in a disentangled manner resulting in more precise variance control.
In experiments, VarianceFlow shows superior performance over other state-of-the-art TTS models both in terms of speech quality and controllability.

%% file: 2_introduction.tex
\section{Introduction}
\label{sec:intro}
In Text-to-Speech (TTS), the one-to-many relationship between text and speech is one of the major problems that makes it challenging to learn the text-to-speech conversion.
The early autoregressive (AR) TTS models \cite{shen2018natural, li2019neural} dealt with the difficulty by factorizing the speech distribution into the product of homogeneous conditional factors in sequential order.
However, although they succeeded in generating high-quality speech, their slow inference speed and exposure bias were inevitable problems inherent in the AR models.
  
As it has advanced from AR TTS models to non-AR TTS models, two types of methods for the non-AR models to solve the one-to-many problem have been proposed.
The first type (Type-$\RN{1}$) is to use an advanced generative framework such as normalizing flow \cite{NEURIPS2020_5c3b99e8}, diffusion model \cite{pmlr-v139-popov21a}, and generative adversarial network \cite{yang21e_interspeech}.
Unlike the mean squared error (MSE) based training that assumes a Gaussian distribution, these frameworks do not assume any pre-defined distribution for a target distribution.
As a result, they can generate high-quality and diverse speech samples compared to the previous non-AR TTS models trained with the MSE loss \cite{NEURIPS2019_f63f65b5, Peng2020nonautoregressive}.
The second type of the methods (Type-$\RN{2}$) is to solve a task of TTS by dividing it into simpler two tasks based on variance information such as pitch or energy \cite{ren2021fastspeech, lancucki2021fastpitch, bak21_interspeech}: (1) text conditioned variance modeling; (2) text and variance information conditioned speech generation.
Then, by using ground-truth variance information when learning the two tasks, Type-$\RN{2}$ models achieve faster training convergence and higher speech quality.
Moreover, unlike Type-$\RN{1}$ models, Type-$\RN{2}$ models can explicitly control the variance factors by manipulating the variance values used in the speech generation.

\begin{figure}[t]
\includegraphics[width=8.6cm]{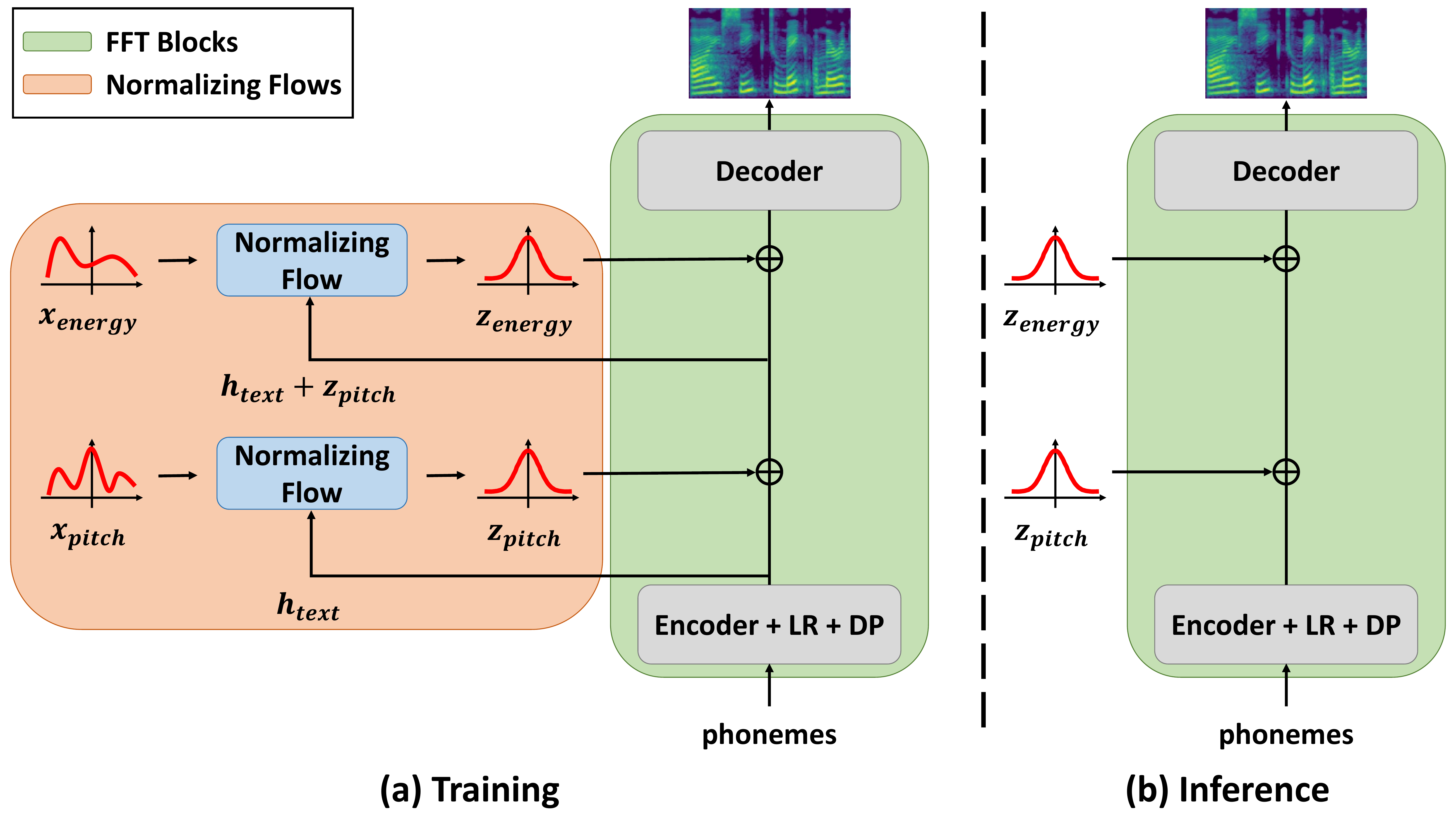}
\centering
\caption{The architecture of VarianceFlow. LR denotes a length regulator that expands text representations to the length of a melspectrogram based on phoneme durations. DP denotes a duration predictor that predicts the phoneme durations.
}
\vspace{-0.2cm}
\label{fig: varianceflow}
\end{figure}

In this paper, we propose VarianceFlow, a novel model that combines the advantages of Type-$\RN{1}$ and Type-$\RN{2}$, achieving a high-quality and controllable TTS model.
Unlike the previous Type-$\RN{2}$ models that learn the text conditioned variance modeling based on the MSE loss, our model uses normalizing flow (NF) for the variance modeling (Figure \ref{fig: varianceflow}).

There are two large advantages of using NF for the variance modeling.
First, since NF is robust to the one-to-many problem, it learns the variance distribution better than using the MSE loss \cite{DBLP:conf/iclr/DinhSB17}, and it leads to improved speech quality.
Second, NF enhances the variance controllability by disentangling the latent variance representation and the text \cite{NEURIPS2020_1cfa81af}.

In experiments, VarianceFlow shows its superiority in variance modeling by showing better speech quality compared to other AR and non-AR state-of-the-art TTS models.
In addition, VarianceFlow shows its more precise variance controllability by showing that it uses the provided pitch values more intactly.
Lastly, we show that VarianceFlow can generate diverse speech samples given a text input.

%% file: 4_methodology.tex
\section{VarianceFlow}
In this section, we first explain FastSpeech 2 \cite{ren2021fastspeech}, which is a baseline of our model.
Then, we present VarianceFlow that learns to match a variance distribution to a prior distribution based on normalizing flow.
\subsection{FastSpeech 2} \label{sec: fastspeech 2}
FastSpeech 2 is a Type-$\RN{2}$ non-AR TTS model that uses pitch and energy as additional variance information in speech generation\footnote{Phoneme durations are also used as variance information, but it is out of focus of this work.}.
Firstly, a phoneme sequence is encoded by an FFT encoder, and it is expanded to the length of the target melspectrogram given phoneme durations.
Next, ground-truth pitch and energy values extracted in frame-level are projected to the space of the text representation.
Lastly, the projected variance vectors are added to the expanded text representation, and an FFT decoder generates a melspectrogram based on it.
  
Since the ground-truth variance information is unavailable at inference, FastSpeech 2 has additional modules called variance predictor for variance modeling.
During training, the variance predictors are trained with the MSE loss to predict the ground-truth variance values based on the input text.
Then, the variance values predicted by the variance predictors are used at inference.
However, there is still a problem that predicting pitch or energy from a text is also a difficult task having a one-to-many relationship.
For this, FastSpeech 2 trains its pitch predictor with wavelet transformed pitch spectrogram instead of raw pitch values.
Also, FastPitch \cite{lancucki2021fastpitch}, which uses pitch information similar to FastSpeech 2, uses phoneme-averaged pitch values to ease the difficulty of learning the complex pitch distribution.

\subsection{VarianceFlow}
VarianceFlow is a model that uses variance information such as pitch and energy based on normalizing flow \cite{DBLP:journals/corr/DinhKB14, DBLP:conf/iclr/DinhSB17}.
During training, pitch and energy variance information is provided to a feed-forward transformer (FFT) decoder via NF modules (Figure \ref{fig: varianceflow}).
Then, VarianceFlow learns to generate speech based on the information and the input text.
Meanwhile, the NF modules learn to match a latent variance distribution to a simple prior distribution.
At inference, the variance information is provided to the FFT blocks by directly sampling the latent representation from the prior.

A normalizing flow (NF) consisting of consecutive bijective transforms converts a complex variance distribution to a simple prior distribution.
Based on the change of variables formula, using bijective transforms enables to calculate probability density of the variance information by setting the latent variance distribution as a simple prior distribution:
\begin{equation} \label{eq:1}
\log \boldsymbol{p_{\theta}(x|h)} = \log \boldsymbol{p(z)} + \sum_{i=1}^k \log|\det(\boldsymbol{J}(\boldsymbol{f_i}(\boldsymbol{x;h})))|
\end{equation}
\begin{equation} \label{eq:2}
\boldsymbol{z}= \boldsymbol{f_k} \circ  \boldsymbol{f_{k-1}} \circ \ldots \circ \boldsymbol{f_0(x;h)},
\end{equation}
where $\boldsymbol{x}$ is a variance factor, $\boldsymbol{h}$ is a hidden representation, $\textbf{z}$ is a latent representation, and $\boldsymbol{f_i}$ is a bijective transform.
Therefore, using a unit Gaussian distribution as the prior distribution, VarianceFlow is trained to maximize the log-likelihood of the variance information based on (\ref{eq:1}) and (\ref{eq:2}).
To fully utilize parallel computation with enough expressive power, VarianceFlow uses rational-quadratic coupling transform \cite{NEURIPS2019_7ac71d43}.
  
\subsection{Loss function}
Replacing the MSE losses for pitch and energy predictors in FastSpeech 2 \cite{ren2021fastspeech} with the NF losses of VarianceFlow, the final objective function of VarianceFlow becomes as follows:
\begin{equation} \label{eq:3}
\mathcal{L}_{total} = \mathcal{L}_{melspec} + \mathcal{L}_{duration} + \alpha \cdot \mathcal{L}_{pitch} + \alpha \cdot \mathcal{L}_{energy},
\end{equation}
where the first two terms are the losses existing in FastSpeech 2, and the last two terms are the negative log-likelihood NF losses for pitch and energy.
Here, each of the NF losses ($\mathcal{L}_{NF}$)  not only learns the variance modeling but also disentangles the variance factor with the text, and it is shown by the fact that $\mathcal{L}_{NF}$ can be decomposed as follows \cite{NEURIPS2020_1cfa81af}:
\begin{equation} \label{eq:4}
\mathcal{L}_{NF} = D_{KL}\left[\boldsymbol{q_{\theta}(z|h)} \parallel \boldsymbol{p(z)}\right] + H(\boldsymbol{x|h})
\end{equation}
For an NF module in VarianceFlow, the second entropy term is constant, so it is trained to minimize the kullback-leibler divergence between the conditional latent variance distribution $\boldsymbol{q_{\theta}(z|h)}$ and the prior distribution $\boldsymbol{p(z)}$.
It means that it is trained to make $\boldsymbol{z}$ and $\boldsymbol{h}$ disentangled because the prior $\boldsymbol{p(z)}$ is chosen independently with $\boldsymbol{h}$.
As a result, it enhances the responsiveness of the model to $\boldsymbol{z}$, resulting in more precise control.
  
\subsection{Controlling a variance factor}
\begin{figure}[ht]
\includegraphics[width=7cm]{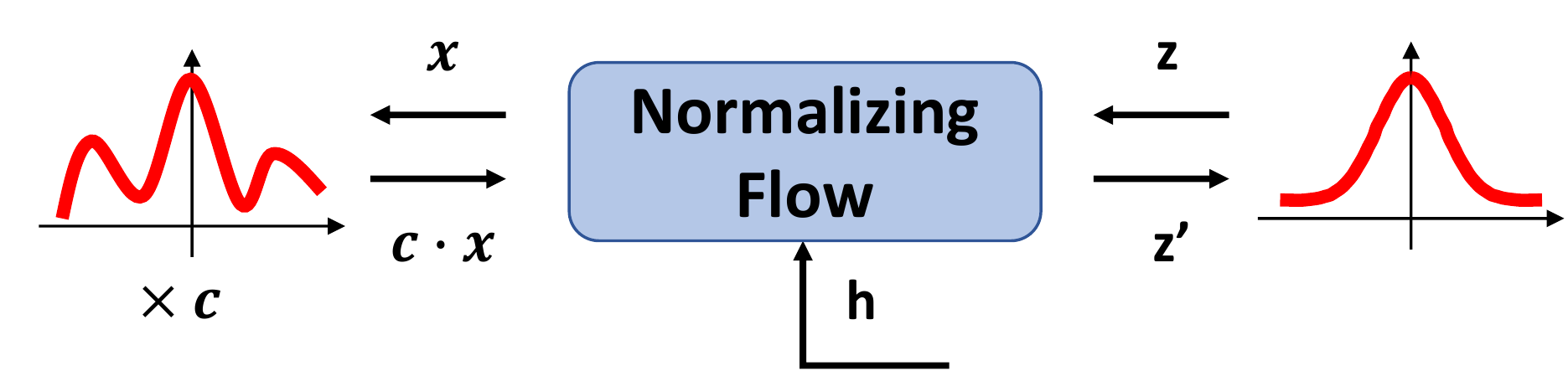}
\centering
\caption{Flow of signals when controlling a variance factor}
\label{fig:control}
\end{figure}
The invertibility of NF makes it possible to obtain the raw variance value $\boldsymbol{x}$ from the latent representation $\boldsymbol{z}$.
Therefore, it is also possible for VarianceFlow to control the variance factors as other Type-$\RN{2}$ TTS models (Figure \ref{fig:control}).
To control the variance information, the sampled latent representations are first brought to the raw variance space using the inverse transforms of NF.
Then, the variance factor is manipulated in the space (\emph{e.g.}, multiplying a constant to the raw variance value $\boldsymbol{x}$).
Lastly, the manipulated value is provided to the FFT decoder via NF again.

%% file: 5_experiments.tex
\section{Experiments}
In this section, experimental setup and results are provided.
Also, the audio samples and more experimental figures are available at \url{https://leeyoonhyung.github.io/VarianceFlow-demo}.

\subsection{Experimental setup}
In experiments, LJSpeech dataset \cite{ljspeech17} is used.
The waves are converted to log-melspectrograms with 1024 fft window and 256 hop lengths.
The pitch values are extracted using Parselmouth \cite{parselmouth} and they are converted to log-scale and pitch spectrogram is not used.
When calculating the total loss of VarianceFlow, $\alpha=0.1$ is used for the coefficient of the NF losses.
When generating audio samples, we use a zero-mean Gaussian prior distribution with a standard deviation $\sigma=0.333$, and we use HiFi-GAN vocoder \cite{NEURIPS2020_c5d73680}.
As a baseline, we adopt FastSpecch 2 architecture \cite{ren2021fastspeech} and also adopt its FFT architecture for VarianceFlow.
For an NF module of VarianceFlow, we use 4-layer rational-quadratic coupling transform architecture following the architecture of the stochastic duration predictor in \cite{pmlr-v139-kim21f}.
We train all models for 230k steps with a batch size of 16.
We use AdamW optimizer \cite{loshchilov2018decoupled} with $\beta_1=0.9$ and $\beta_2=0.98$ with the Noam learning rate scheduling \cite{NIPS2017_3f5ee243}.

\subsection{Speech quality} \label{subsec: speech quality}
To see the effectiveness of VarianceFlow in improving speech quality, we compare the speech quality of VarainceFlow with FastSpeech 2 based on the 9-scale mean opinion score (MOS), while varying the ways of using the variance information either phoneme-averaged level or frame level (Table \ref{tab: mos}).
For each model, fifty samples from the test set are used, and five testers living in the U.S. are asked to give a score ranging from 1 to 5 to each audio sample.
For FastSpeech 2, MOS is better when variance information is used in the phoneme-averaged level.
It means that training the current architecture of the variance predictors using the MSE loss lacks in learning the complex variance distribution.
On the contrary, VarianceFlow shows better speech quality when the variance information is provided in the frame level.
It means that our VarianceFlow learns the complex variance distribution well based on NF.
Although the architectures used in variance modeling are different, directly using the latent representation sampled from a Gaussian distribution does not slow down the synthesis speed or increase the number of parameters at inference.
Overall, VarianceFlow trained with the frame level variance information shows the best speech quality including other state-of-the-art AR and non-AR TTS models, Tacotron 2 and Glow-TTS.
\input{tables/mos}

\subsection{Controllability}
\input{tables/pitch_control}
To compare the controllability of the models, we conduct objective and subjective evaluations while adjusting the predicted pitch values with various pitch shift coefficients $\lambda$.
In addition, to see the effectiveness of the NF loss in disentangling the variance factors with other factors, we also train a model by reverting the direction of NF (`VarianceFlow-reversed').
For this model, the raw variance information is directly provided to the FFT decoder while the NF modules separately learn to match the latent variance distribution and the prior distribution.
In this case, it still has the advantage of using advanced distribution modeling method, but it uses the variance information without disentangling.

\subsubsection{Pitch responsiveness}
\begin{figure}[ht]
\includegraphics[width=8.6cm]{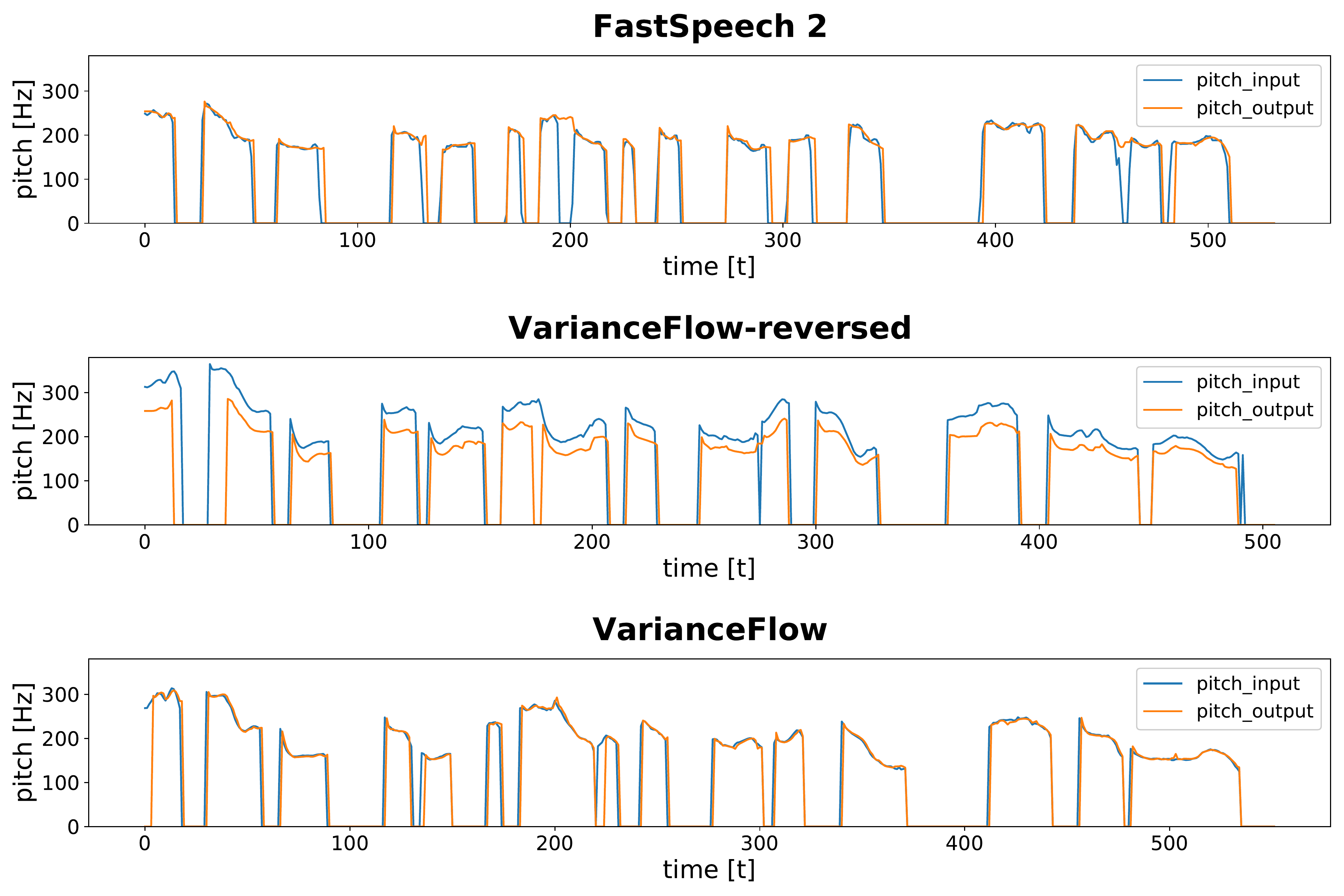}
\centering
\caption{Each figure shows f0 contours of input pitch and extracted pitch when $\lambda=0$.}
\label{fig: pitch_contour_comparison}
\vspace{-0.3cm}
\end{figure}
To evaluate the responsiveness of the models to the variance information, we measure f0 frame error rate (FFE) \cite{chu2009reducing} between the pitch values provided to the decoder and the pitch values extracted from the generated audio samples.
All audio samples are generated by adjusting the pitch values in a semitone unit as follows \cite{bak21_interspeech}:
 \[f_{\lambda}=2^{\frac{\lambda}{12}} \times f_0\]
where $f_0$ is the pitch value before shifted.
In the experiments, $\lambda=\{-6, -4, -2, +2, +4, +6\}$ are used.
As shown in Table \ref{tab: pitch control} and Figure \ref{fig: pitch_contour_comparison}, VarianceFlow uses the provided pitch most intactly for most $\lambda$.
Therefore, it is possible to control the pitch with VarianceFlow precisely.
VarianceFlow-reversed shows the largest FFE, and it can be understood based on Figure \ref{fig: pitch_contour_comparison}.
Figure \ref{fig: pitch_contour_comparison} shows that VarianceFlow-reversed learns to use the variation of provided pitch, but it ignores the average of the provided pitch when generating a speech.

\subsubsection{Speech quality with pitch shift}
To evaluate the models subjectively, we measure the MOS for the samples generated using the shifted pitch values in the same way with Section \ref{subsec: speech quality}.
For each model and $\lambda$, twenty samples from the test set are used.
The results in Table \ref{tab: pitch control} show that our VarianceFlow achieves better speech quality for all $\lambda$ compared to FastSpeech 2, and even outperforms VarianceFlow-reversed for $\lambda=\{-2, +2\}$.
We conjecture that this is because while the latent representation is passed to the FFT decoder, NF loss works as a regularization.
Although VarianceFlow becomes worse than VarianceFlow-reversed when $\lambda$ gets larger, we attribute this to the fact that VarianceFlow-reversed does not follow the pitch shift.

\subsection{Diversity}
\begin{figure}[t]
\includegraphics[width=8.6cm]{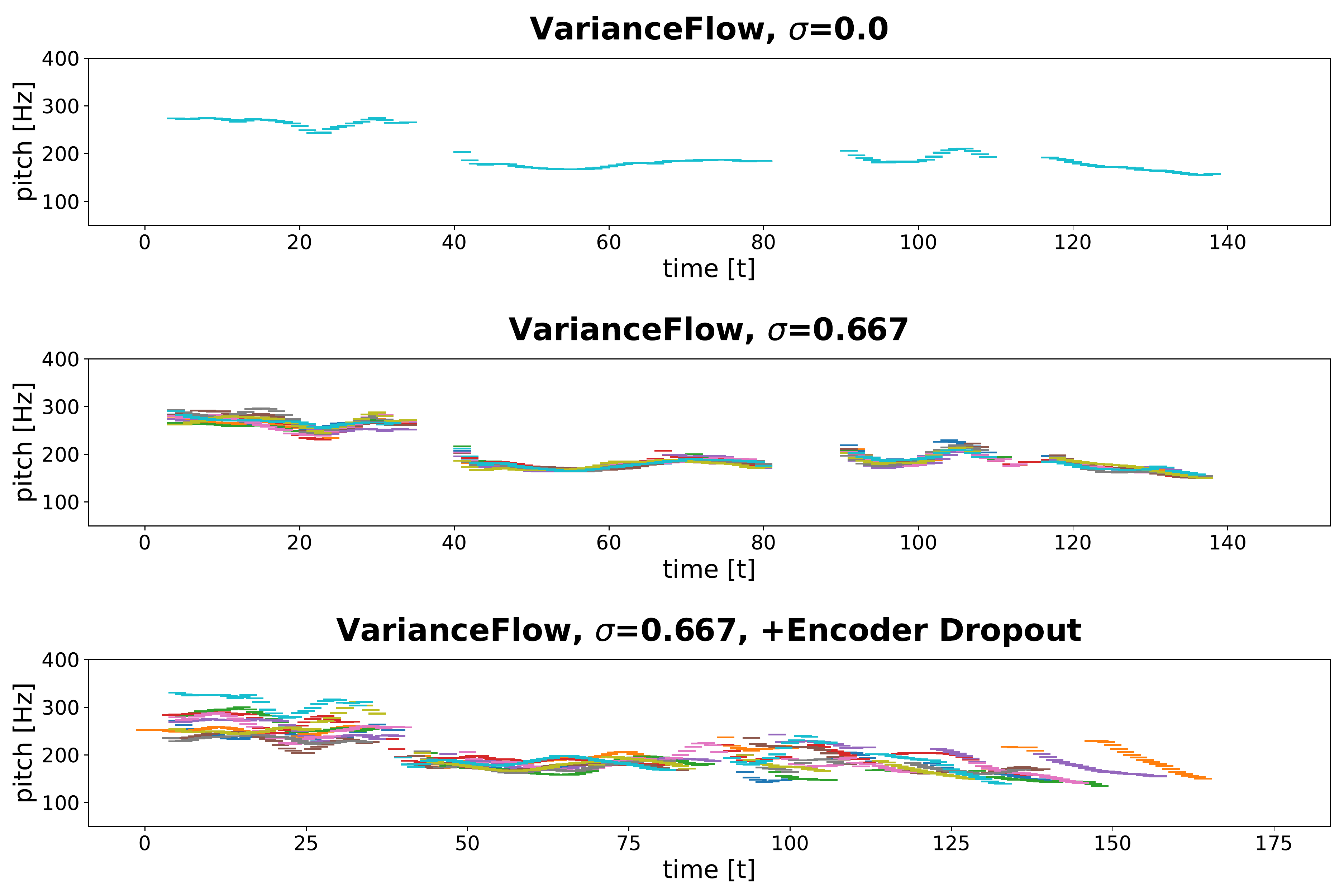}
\centering
\caption{Each figure shows f0 contours of 10 different samples generated using the same $\sigma$.}
\label{fig: f0contour}
\vspace{-0.3cm}
\end{figure}
To see how the variation in latent space affects the variance factors, we generate diverse speech samples from a single text while using different standard deviation $\sigma=\{0.0, 0.667\}$ for latent sampling (Figure \ref{fig: f0contour}).
Here, the $\sigma$ values are chosen in a range where the speech quality is maintained.
In the experiment, it is shown that VarianceFlow can generate natural speech without using latent representations, and pitch variance in samples appears when using the latent representations.
However, the different samples have similar prosody and perceptually sound similar.
We conjecture that this is because it is trained to minimize the conditional entropy $H(\boldsymbol{x} | \boldsymbol{h})$ (Eq. (\ref{eq:4})).
Therefore, we repeat the samplings while having the dropout layers in the FFT encoder activated \cite{pmlr-v139-kim21f, valle2021flowtron}.
In this case, the diversity in terms of prosody increases with varying durations without much loss of speech quality.

%% file: tables/mos.tex
\begin{table}[t]
    \caption{9-scale MOS results with 95\% confidence intervals.}
    \vspace{0.5cm}
	\centering
	\begin{tabular}{ l | c}
		\toprule
		Model & MOS\\
		\midrule
		GT Waveform & 4.47 $\pm$ 0.07\\
		GT Melspectrogram & 4.34 $\pm$ 0.08\\
		\midrule
		Tacotron 2 & 4.03 $\pm$ 0.07\\
		Glow-TTS & 3.72 $\pm$ 0.13\\
		FastSpeech 2-phoneme & 3.92 $\pm$ 0.07\\
		FastSpeech 2-frame & 3.66 $\pm$ 0.09\\
		VarianceFlow-phoneme & 4.04 $\pm$ 0.08\\
		VarianceFlow-frame & \textbf{4.19 $\pm$ 0.07}\\
		\bottomrule
	\end{tabular}
	\label{tab: mos}
	\vspace{-0.25cm}
\end{table}

%% file: tables/pitch_control.tex
\begin{table*}[t]
    \caption{FFE (\%) and 9-scale MOS results measured with different pitch shift scale $\lambda$.}
	\centering
	\renewcommand{\tabcolsep}{1.8mm}
	\begin{tabular}{ l | c c | c c | c c | c c | c c | c c }
		\toprule
		\multirow{2}{*}{Model} & \multicolumn{2}{c}{$\boldsymbol{\lambda=-6}$} & \multicolumn{2}{c}{$\boldsymbol{\lambda=-4}$} & \multicolumn{2}{c}{$\boldsymbol{\lambda=-2}$} & \multicolumn{2}{c}{$\boldsymbol{\lambda=+2}$} & \multicolumn{2}{c}{$\boldsymbol{\lambda=+4}$} & \multicolumn{2}{c}{$\boldsymbol{\lambda=+6}$}\\
		& FFE & MOS &  FFE & MOS &  FFE & MOS &  FFE & MOS &  FFE & MOS &  FFE & MOS\\
		\midrule
		FastSpeech 2 & \textbf{15.94} & 2.93 & 14.00 & 3.46 & 12.61 & 3.65 & 10.94 & 3.29 & 11.57 & 2.63 & 12.24 & 2.04\\
		VarianceFlow-reversed & 21.93 & \textbf{3.77} & 35.97 & \textbf{4.01} & 53.47 & 4.00 & 66.37 & 3.90 & 67.07 & \textbf{3.69} & 67.03 & \textbf{3.53}\\
		VarianceFlow & 24.14 & 3.43 & \textbf{12.16} & 3.87 & \textbf{9.02} & \textbf{4.05} & \textbf{7.26} & \textbf{3.95} & \textbf{7.52} & 3.39 & \textbf{7.68} & 2.74\\
		\bottomrule
	\end{tabular}
	\label{tab: pitch control}
	\vspace{-0.3cm}
\end{table*}

%% file: 6_conclusion.tex
\section{Conclusion}
We propose VarianceFlow, a novel model that uses variance information based on normalizing flow (NF).
VarianceFlow shows that it learns the variance distribution better with improved speech quality.
Also, the invertibility of NF maintains its ability to control the variance factors, and we show that the objective of NF even enhances the variance controllability.

%% file: 7_acknowledge.tex
\section{Acknowledgments}
K. Jung is with ASRI, Seoul National University, Korea.
This work was supported by the National Research Foundation of Korea (NRF) grant funded by the Korea government (No. 2021R1A2C2008855).
This research resulted from a study on the ``HPC Support'' Project, supported by the `Ministry of Science and ICT' and NIPA.